# RELIANCE: Reliable Ensemble Learning for Information and News Credibility Evaluation

**Majid Ramezani**
*Faculty of Computer Science and Information Technology, Institute for Advanced Studies in Basic Sciences (IASBS), Zanjan, Iran.*
*ramezani@iasbs.ac.ir*
ORCID: 0000-0003-0886-7023

**Hamed Mohammad-Shahi**
*Faculty of Computer Science and Information Technology, Institute for Advanced Studies in Basic Sciences (IASBS), Zanjan, Iran.*
*hamedshahi@iasbs.ac.ir*
ORCID: 0009-0008-7405-2192

**Mahshid Daliry**
*Faculty of Computer Science and Information Technology, Institute for Advanced Studies in Basic Sciences (IASBS), Zanjan, Iran.*
*mahshiddaliry@iasbs.ac.ir*
ORCID: 0009-0006-0449-0950

**Soroor Rahmani**
*Faculty of Computer Science and Information Technology, Institute for Advanced Studies in Basic Sciences (IASBS), Zanjan, Iran.*
*soroorrahmani@iasbs.ac.ir*
ORCID: 0009-0000-7465-3766

**Amirhossein Asghari**
*Faculty of Computer Science and Information Technology, Institute for Advanced Studies in Basic Sciences (IASBS), Zanjan, Iran.*
*amirasghari@iasbs.ac.ir*
ORCID: 0009-0003-1000-8233

*Abstract*—In the era of information proliferation, discerning the credibility of news content poses an ever-growing challenge. This paper introduces RELIANCE, a pioneering ensemble learning system designed for robust information and fake news credibility evaluation. Comprising five diverse base models, including Support Vector Machine (SVM), naïve Bayes, logistic regression, random forest, and Bidirectional Long Short Term Memory Networks (BiLSTMs), RELIANCE employs an innovative approach to integrate their strengths, harnessing the collective intelligence of the ensemble for enhanced accuracy. Experiments demonstrate the superiority of RELIANCE over individual models, indicating its efficacy in distinguishing between credible and non-credible information sources. RELIANCE, also surpasses baseline models in information and news credibility assessment, establishing itself as an effective solution for evaluating the reliability of information sources.

## I. Introduction

In the era of information abundance, discerning the reliability of news documents has become a paramount challenge. The rapid dissemination of news through various online platforms has created a fertile ground for misinformation, disinformation, and fake news. As society increasingly relies on digital sources for information consumption, ensuring the credibility of news content becomes imperative.

In the face of global crises such as the COVID-19 pandemic, natural disasters like earthquakes, and geopolitical conflicts including wars, the reliability assessment of news becomes not merely a scholarly pursuit but a critical imperative for societal well-being. Accurate and timely information plays an indispensable role in crisis management, public safety, and policy formulation. The spread of misinformation during such events can have profound and far-reaching consequences, ranging from public panic and misguided responses to compromised public health efforts. In the case of a pandemic like COVID-19, misinformation about the virus, its origins, or potential treatments can undermine public trust in health authorities and exacerbate the challenges of containment. Similarly, during natural disasters or armed conflicts, misleading information can hinder effective evacuation procedures, relief efforts, and diplomatic initiatives. Thus, the ability to distinguish between reliable and unreliable news in these contexts is paramount, underscoring the crucial need for advanced and robust fake news detection system. This field of study is known by various names, including *fake news detection* [1-2], *rumor detection* [3-4], *misinformation detection* [5-6], *information credibility evaluation or assessment* [7-8], *trustworthiness assessment*, and more. Its aim is to assess the reliability or authenticity of information and news documents [9].

The proliferation of misinformation poses a threat to public discourse, decision-making processes, and societal trust in media. Traditional methods of news verification struggle to keep pace with the evolving landscape of deceptive tactics employed by malicious actors. In this context, our paper introduces **RELIANCE (R**eliable **E**nsemble **L**earning for **I**nformation **a**nd **N**ews **C**redibility **E**valuation**)**, an innovative approach leveraging reliable ensemble learning for robust information and news credibility evaluation. It seeks to address these challenges by harnessing the power of ensemble learning, combining the strengths of multiple base models to enhance the reliability and accuracy of news credibility assessment. To do so we have proposed five base models for credibility evaluation, including *Support Vector Machine (SVM)-based*, *naïve Bayes-based, logistic regression-based, random forest-*

*based, and Bidirectional Long Short Term Memory Networks (BiLSTMs)-based* models. Together, these independent base methods collaborate to minimize the generalization error in predictions through ensemble learning. In essence, ensemble learning involves employing multiple diverse base models to predict outcomes, surpassing the predictive capabilities of individual models [10]. Therefore, the authors proposed RELIANCE which stands as a pioneering effort to combine the capabilities of individual classifiers, thereby creating a more resilient and trustworthy system for distinguishing between credible and non-credible information and news documents.

The current study provides *two significant contributions* to distinguishing the credibility and non-credibility of information and news documents: introducing five distinct diverse methods for information news credibility evaluation, and enhancing credibility evaluation accuracy through the ensemble learning of the aforementioned models as the base models during ensemble learning.

The structure of our study is as follows: the *literature review* comprehensively surveys existing research. The *methods* details our innovative approach, including five distinct credibility evaluators as well as ensembling them through RELIANCE. Subsequently in *results and discussion*, we present and analyze the *results* obtained from our experiments, assessing the proposed model's performance. This section also critically interprets the findings, highlighting the contributions and implications of our work. Finally, the *conclusion* highlights the key insights, reinforces the significance of our proposed methodology.

II. LITERATURE REVIEW

The rapid dissemination of information through digital platforms has directed in an era where the credibility of news content is frequently challenged. In response to the proliferation of misinformation and fake news, researchers have proposed different approaches to evaluate the credibility of news documents. Throughout the research history of our work, this field has been recognized by various names, including *rumor detection*, *fake news detection*, and *reality detection*. Perusing the literature, one can categorize the previous studies into three categories: **text-based methods** [11-18], **context-based methods** [15, 19-23], and **hybrid methods** [24-28]. Table 1 provides an analytic perspective on related works in information and news credibility evaluation in a tabular form.

The **text-based methods** primarily rely on miscellaneous information that can be acquired from the text documents. In a basic investigation outlined in [11], the authors introduced a method that combines lexical features, word embeddings, and n-gram characteristics for identifying the stance in fake news. The researchers in [12] introduced a new computational method for automatically detecting fake news, relying on a novel text analysis approach centered around the linguistic features provided by LIWC (Linguistic Inquiry and Word Count) [16]. In another investigation, the authors to perform news credibility evaluation have introduce a hybrid model based on Logistic Regression and n-gram analysis [13]. They have used Term Frequency-Inverted Document Frequency (TF-IDF) as feature extraction technique. They have also suggested a Linear Support Vector Machine (LSVM) model to perform news credibility evaluation. The authors in [14] introduced a model that integrates several key features of text document (such as the text of news document, news source, etc.) for a more precise and automated prediction. Guided by these features, they proposed a model named CSI, comprising three modules: Capture, Score, and Integrate. The initial module relies on responses and text, utilizing a Recurrent Neural Network to capture the temporal patterns of user activity on a given news document. The second module understands news source characteristics based on user behavior, and these are integrated with the third module to classify a news document as either reliable or unreliable.

The **context-based methods** basically attempt to utilize those information that are provided by the context such as user-based features, or post-based features. The authors in [15, 19] have proposed models that user-specific characteristics are derived from individual profiles to assess their attributes for the sake of content credibility evaluation. More over in several investigations to perform credibility evaluation, features based on network structures are primarily extracted through the development of precise detection systems [20-21]. This involves utilizing diffusion networks, association networks, and propagation networks to assess credibility of text documents. Given the extensive integration of social media, research also incorporates social media interactions in the detection of fake news. For instance, there is a focus on early detection through social learning and user-based relationships [22-23].

The **hybrid methods** try to consider all of the available information to evaluate the credibility of the text documents. The authors in [24] presented an innovative hybrid system for detecting fake news, which integrates linguistic and knowledge-based methodologies, combining their respective strengths. Their hybrid approach employs two distinct feature sets: linguistic features (such as title, word count, reading ease, lexical diversity, and sentiment), and a new set of knowledge-based features called fact-verification features. The authors in [25] introduced a novel transformer-based model named X-CapsNet, incorporating Convolutional Neural Networks or CNN (CapsNet). X-CapsNet employs a CapsNet with a dynamic routing algorithm and integrates a size-based classifier for distinguishing between short and long fake news statements. The detection of long fake news statements is facilitated by a Deep CNN, while short news statements are identified using a Multi-Layer Perceptron (MLP). Considering the context of social networks the authors in [26] have presented a novel approach to identify fake news by analyzing the temporal propagation tree. They continuously extract relevant features from the constructed propagation trees and employ a type of recurrent neural network (LSTM) to capture the temporal dynamics of these features. This enables them to discern the evolving pattern of the propagation tree over time, aiding in the estimation of the credibility of a news article. In their seminal study, the authors in [27] have explored the feasibility of unsupervised fake news detection by considering the truths of news and users' credibility as latent random variables. Utilizing users' engagements on social media, they have discerned their opinions regarding the authenticity of news.

In current study, to leverage different credibility evaluation models abilities, we proposed an ensemble learning model to perform information and news credibility evaluation, called RELIANCE. The combination of diverse models through ensemble learning offers a synergistic approach, harnessing the complementary strengths of individual methods, thereby enhancing the robustness and accuracy of news credibility evaluation compared to the performance of individual models.

## III. METHODS

The objective of our investigation was to improve the automated evaluation of information and news credibility by employing ensemble learning. Ensemble learning, combining various methods, enhances predictive performance beyond that of individual methods. Consequently, we introduced five distinct base models, each operating at different processing levels, for integration into the ensemble learning model. Namely *Support Vector Machine (SVM)-based, naïve Bayes-based, logistic regression-based, random forest-based, and Bidirectional Long Short-Term Memory Networks (BiLSTMs)-based models*. To this end, we proposed **RELIANCE (Reliable Ensemble Learning for Information and News Credibility Evaluation)**, that performs credibility evaluation in three phases; *Phase 1* focuses on preprocessing the input news text documents to prepare them for the main process. *Phase 2* performs feature engineering through embedding the input news text documents. *Phase 3* is devoted to evaluate the credibility of news text documents. The general architecture of RELIANCE is depicted in Fig. 1.

### A. Phase 1: Preprocessing

This conventionally well-established and significant stage within the domain of natural language processing aims to prepare the input text and convert it into a more easily comprehensible format for machines during the main process. The preprocessing activities involved may vary depending on the task at hand. The applied preprocessing activities in RELIANCE as can be seen in Fig. 1 are as listed below.

● **Tokenization**: Tokenization is the procedure of dividing a text into segments known as tokens, which are typically associated with words. For this purpose, we used the Natural Language Toolkit (NLTK) [31].

● **Noise Removal:** To enrich the quality of the input text, we performed noise removal steps including *stop words removal*, and *punctuation* and *signs removal* since they do not convey any specific information that is essential to the performance of the system. Similar, to the previous step, NLTK was also used to perform stop words removal.

TABLE I. AN ANALYTIC VIEW ON RELATED WORKS IN INFORMATION AND NEWS CREDIBILITY EVALUATION IN A TABULAR FORM

| Authors | Category | Dataset | Method | Input Genre |
|---------|----------|---------|--------|-------------|
| Ghosh & Shah[29] | Text-based | BuzzFace | Commination of techniques from information retrieval, natural language processing, and deep learning | News Documents |
| Liu & Wu [28] | Hybrid | Weibo, Twitter15, and Twitter16 | Time series classifier with both recurrent and convolutional networks | Social media content |
| Seddari et al. [12] | Hybrid | Buzzfeed Political News dataset | Random Forest, Logistic Regression, Additional Trees Discriminant, and eXtreme Gradient Boosting (XGBoost) | News Documents |
| Kaliyar et al.[17] | Text-based | FN-COV | A combination of convolutional neural network and long short term memory network | COVID-19 |
| Goldani et al. [25] | Hybrid | Liar datasets | A novel transformer-based model named X-CapsNet | COVID-19 |
| Dixit et al. [18] | Text-based | LIAR, LIAR-PLUS, and ISOT | Fuzzy Convolutional Recurrent Neural Network (CRNN) | News documents |
| Davoudi et al. [26] | Hybrid | PolitiFact and GossipCop | Long Short Term Memory networks (LSTMs) | News documents |
| Yang et al. [27] | Hybrid | LIAR and BuzzFeed News | Bayesian network | Social media |
| Basak et al. [15] | Context-based | Shaming detection of tweets | Support vector machine | Twitter |
| Potthast et al. [19] | Context-based | BuzzFeed-Webis Fake News | Similarity detection between text categories | News Documents |
| Karimi et al. [20] | Context-based | FakeNewsNet | Multi-source Multi-class Fake news Detection framework MMFD | News Documents |
| Gupta et al. [21] | Context-based | FakeNewsNet | A tensor factorization | News Documents |
| Nguyen et al. [22] | Context-based | Stance-annotated dataset | Graph Convolutional Networks (GCN) | News Documents |
| Shu et al. [23] | Context-based | FakeNewsNet | TriFN (which models publisher-news relations and user-news interactions) | News Documents |
| Ghanem et al. [11] | Text-based | Fake NewsChallenge (FNC-1) | Combination of lexical features, word embeddings, and n-gram statistics | News documents |
| Singh et al. [12] | Text-based | Kaggle FakeNews | linguistic features provided by LIWC | News documents |
| Ahmed et al. [13] | Text-based | Kaggle FakeNews | Logistic Regression and n-gram statistics | News documents |
| Ruchansky et al.[14] | Text-based | Kaggle FakeNews | Recurrent Neural Network and a set of text-based features | News documents |
| Yang et al. [30] | Text-based | Kaggle FakeNews | Latent feature extraction using convolutional neural networks | News documents |

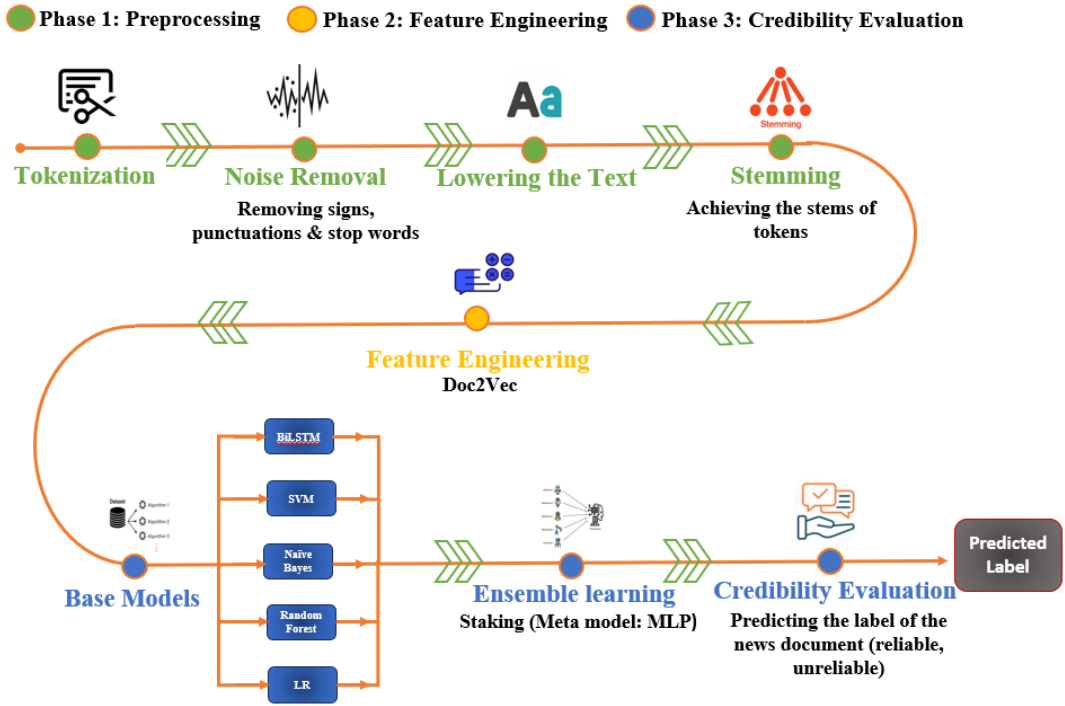

Fig. 1. The general architecture of the proposed method (RELIANCE)

● **Lowering the Text:** The process of lowering the text uniforms all the characters and reduces the variability in text and makes it more consistent.

● **Stemming:** Is the morphological analysis of words that reduces their inflected forms to their stem [32]. This process is achieved by removing affixes from words. Escalating the performance of the natural language processing tasks in the consequence of stemming. The NLTK was also used for this process.

*B. Phase 2: Feature Engineering*

Feature engineering plays a pivotal role in the realm of natural language processing, encompassing the identification, transformation, and extraction of pertinent features from raw textual data. This method facilitates the determination of the most informative features for primary processing, concurrently eliminating less significant ones. The main goal is to decrease the dimensionality of the dataset by reducing the number of features, generating novel features from the initial ones. Ideally, this refined feature set encapsulates the essential features from the original set.

Doc2Vec [33], an extension of Word2Vec, to transform text documents into vectors within a high-dimensional space. The underlying concept of Doc2Vec is based on the assumption that words appearing together in a document share similar meanings and, consequently, should exhibit similar vector representations. In simpler terms, the vector representation of a word captures its semantic meaning. This approach enables the acquisition of the contextual relationship between words within a document, a critical factor in comprehending the document's overall meaning.

Leveraging Doc2Vec for feature engineering enables the transformation of the input text data into a low-dimension vector space, facilitating processing by machine learning algorithms. In this transformation, the dimensionality reduction effectively eliminates excessive dimensions and retaining only crucial dimensions for subsequent processing.

*C. Phase 3: Credibility Evaluation*

At this point, the data is pre-processed, feature-engineered, and we possess a rich and informative feature set. The proposed method (RELIANCE), performs credibility evaluation of the input documents relying on ensembling several base models. More specifically, RELIANCE leverages the strengths of several credibility evaluation models and combines them to achieve more reliable predictions through stacking. The primary goal of ensemble learning is to increase overall predicting performance by merging the prediction of various models. In reality, it uses the votes from several different base models rather than just one to predict a label.

Among various ensemble techniques, commonly employed ones include *Bagging* (Bootstrap Aggregation), *Boosting*, and *Stacking* [10]. The first two are frequently utilized to integrate homogeneous base models with the goal of reducing variance and minimizing bias, respectively. In contrast, Stacking is typically employed to integrate heterogeneous base models with the aim of enhancing predictive performance. RELIANCE utilizes five credibility evaluators including *(SVM)-based, naïve Bayes-based, LR-based, random forest-based, and BiLSTM-based models* as the base models, each of which, possesses distinctive capabilities tailored to perform credibility evaluation. Hence, considering the heterogeneity of the applied base models, RELIANCE is developed using Stacking. It is significant to point out that RELIANCE utilized a Multi-Layer Perceptron (MLP) as the meta model to produce the final predictions (credibility evaluation).

*1) Dataset*

The Fake News dataset [34] has been used to train and validate all of the models in RELIANCE. This dataset is gathered during the 2016 U.S. Presidential Election.

The dataset comprises two categories of articles: *fake* and *real* news. Real news were gathered by scraping content from Reuters.com, a reputable news website. Fake news were sourced from various unreliable websites identified by Politifact, a U.S.-based fact-checking organization, and Wikipedia. While the dataset encompasses diverse documents types on various subjects, a predominant focus is observed on political and global news topics.

TABLE II. THE PROPERTIES OF FAKE NEWS DATASET (UNEQUAL NUMBERS ARE DUE TO THE ABSENCE OF SOME VALUES IN THE DATASET).

| *Attribute* | *Number of instances in the dataset* |
|---|---|
| ID | 20800 |
| Title | 20242 |
| Author | 18843 |
| Text | 20761 |
| label | 20800 |

Fake News dataset comprises five attributes: ID, Title, Author, Text, and Labels. The ID is a distinctive identifier for each news document. The Title corresponds to the primary heading of the news piece, and the Author indicates the creator's name. The Text is the core part of the dataset, containing the news document's text. Finally, the Label determines the credibility of the news document (assigning fake or real labels). Table II describes the Fake News Dataset's properties. In terms of label distribution, Fake News comprises approximately 51% reliable (real) news documents and approximately 49% unreliable (fake) news, as shown in Fig. 2. As can be understood from the Fig. 2, the dataset is balanced and appropriate for training.

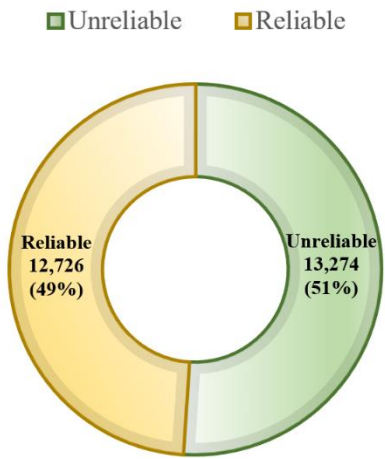

Fig. 2. The distribution of real or fake labels in Fake News dataset

*2) Base Models*

RELIANCE utilizes five distinct credibility evaluation models, including several traditional machine learning models as well as a deep learning model to leverage their capabilities through ensemble learning to ensures the quality of the final predictions.

● **BiLSTM:** we proposed a BiLSTM network to perform credibility evaluation. LSTM Recurrent Neural Networks take into account a series of previous inputs. This characteristic makes LSTMs highly effective at learning from sequential data such as text documents. As a specific type of RNN, LSTM leverages past information to understand each word in the forward direction. Conversely, BiLSTM benefits from considering both past and future words (forward and backward directions) which aligns with how humans understand language. We proposed a 3-layer BiLSTM with respectively 64, 128, and 64 units joined to a dense layer with 64 neurons.

● **Logistic Regression (LR):** we implemented a logistic regression-based model for credibility evaluation of information and news documents. In essence, it is a statistical model that estimates the probabilities of a binary dependent variable. Unlike other methods that operate on independent inputs, LR considers a set of previous inputs, making it particularly effective for learning from sequential data.

● **Support Vector Machine (SVM):** we proposed the SVM-based model, a powerful and versatile machine learning technique used for various tasks including classification, and regression. Unlike other algorithms that operate on independent inputs, SVMs also consider a set of previous inputs and are particularly effective at learning from sequential data. This feature is due to the nature of SVMs, which aims to find the maximum separating hyperplane between different classes in the target label.

● **Random Forest:** we used the Random Forest algorithm, a powerful and widely used machine learning technique for classification and regression tasks. Random Forest creates a set of decision trees and combines their outputs to make the final prediction. Essentially, Random Forest is a combination of several decision trees that individually performs predictions. In practice, ensembling several homogeneous models, Random Forest tries to increase the robustness and stability of model, and reduce the risk of overfitting, and provides more accurate predictions.

● **Naïve Bayes:** The Naive Bayes algorithm is a probabilistic machine learning method primarily utilized for classification tasks. It also considers the prior inputs, making it well-suited for learning from sequential data. The algorithm operates under the assumption that attributes are independent of each other, which is not always the case in reality. Nevertheless, this simplifying assumption surprisingly leads to better predictions. Leveraging the principle of conditional probability, Naive Bayes calculates the probability of an event based on prior knowledge. Despite its simplicity, Naive Bayes demonstrates robust performance and finds widespread use in the field of text classification.

*3) Ensemble Learning*

The fundamental idea behind ensemble learning is to enhance overall predictive performance by aggregating decisions from several distinct base models. It involves considering the votes of various base models to predict the final label, rather than relying on a single model (similar to harnessing the wisdom of the crowd for predictions). By leveraging independent and diverse base models, ensemble learning aims to reduce the generalization error in predictions, providing a compelling incentive for its application.

As discussed earlier, RELIANCE performs ensemble learning through stacking five heterogeneous distinct base models. Specifically, stacking (also known as *stacked generalization*), is a machine learning algorithm that uses a meta-learning algorithm as a *meta-model*, to learn how to combine the best predictions of several base machine learning algorithms. Ensemble learning offers the advantage of utilizing the strengths of multiple models, leading to predictions that outperform any individual model. The aforementioned five base models were chosen as the base credibility evaluation methods, with a MLP selected as the meta-model. Although that the predictions of the base models are not considered as the final predictions, they have undeniable role in final decision making by the RELIANCE. The predictions from the base models, (resembling the wisdom of the crowd), are employed to train the meta-model. Ultimately, the meta-model generates evaluates the credibility of the input documents. To do so, we used a MLP model with three hidden layers, respectively containing 64, 128, and 64 units. As can be seen, Fig. 3 has depicted the architecture of the proposed MLP model.

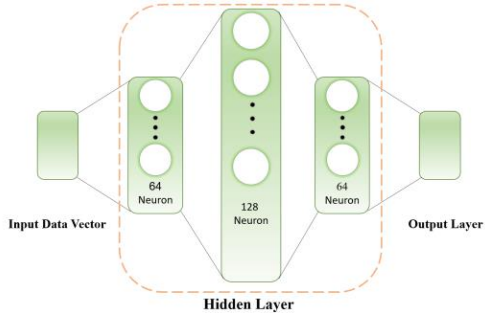

Fig. 3. Architecture of the proposed MLP model

IV. RESULTS AND DISCUSSIONS

*A. Evaluation Metrics*

Traditionally, several of the most common evaluation metrics such as precision, recall, f-measure, and accuracy are used to assess text classification models. In the same manner, we employed all of them, with a particular emphasis on *accuracy* as the primary evaluation metric.

*B. Evaluation Results*

To perform news credibility evaluation, we proposed RELIANCE as an ensemble learning model that combines five distinct classifiers as the base models including *SVM-based, naïve Bayes-based, LR-based, random forest-based, and BiLSTMs-based models*. To perform feature engineering, we employed Doc2Vec with an embedding size equal to 1,200, and the minimum counts of word equal to 1, and conducted the training for 50 epochs. Afterwards, at first, the base models should perform credibility evaluation on the input text documents. Each of the base models adheres to specific settings for generating predictions. The train-test split rate of the data set is 80-20 for all of the models in our investigation. More specifically:

**BiLSTM**: Table III encompasses the applied parameter settings in proposed BiLSTM model.

**Logistic Regression (LR):** For the LR-based model, the maximum number of iterations for the solver (*lbfgs*) to converge is set at 1000. Moreover, the L1 regularization is also used to reduce model generalization error.

**SVM:** For the SVM-based model, the chosen kernel is *'rbf'* (Radial Basis Function). The kernel selection is crucial as it determines the type of decision boundary used for classification. Moreover, the L1 regularization is also used to reduce model generalization error.

**Random Forest:** The Random Forest-based model is configured with the parameter *n_estimators* set to 100. This parameter defines the number of integrating decision trees. The maximum depth of the tree is set to None. That is to say, nodes are expanded until all leaves are pure or until all leaves contain less than minimum number of samples required to split an internal node.

**Naïve Bayes:** We used the Multinomial Naive Bayes algorithm, which is commonly used for text classification. We set *fit_prior* to false. Namely, the uniform prior probabilities are applied. Furthermore, the additive smoothing parameter (alpha) was set to 1.0.

**MLP:** Table IV provides a comprehensive overview of the applied parameter settings in the proposed MLP model. Fig. 4 represents the learning curves of proposed MLP model.

Table V presents the results obtained from the evaluation of five distinct proposed credibility evaluation models as well as the performance of RELIANCE which was our main proposed method the ensembles the base models. It encompasses the values of four evaluation metrics, namely *precision*, *recall*, *F1-score* and *accuracy*. It is noteworthy that, since F1-score neglects all of the correctly false labelled samples by the system (TN), in practice it loses the ground to accuracy while evaluating a credibility evaluation system. Therefore, we will focus on accuracy as the main metric to evaluate the credibility of the models.

As can be seen from Table V, the proposed ensemble learning model, namely RELIANCE that combines the aforementioned base models, has achieved the higher accuracy than individual base models. Moreover, among the base models, SVM-based, LR-based, naïve bayes-based, random forest-based, and BiLSTM-based models respectively achieved the highest accuracies.

In the same manner, the proposed ensemble learning model has achieved higher F1-score rather than individual base models. Similar to accuracy values, among the base models, SVM-based, LR-based, naïve Bayes-based, random forest-based, and BiLSTM-based models respectively achieved the highest values for F1-score. Fig. 5 represents a visual comparison of the performance of all credibility evaluation models proposed in current study.

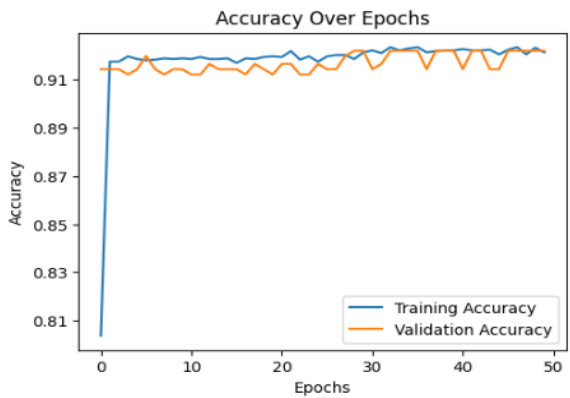

(a) Accuracy curve

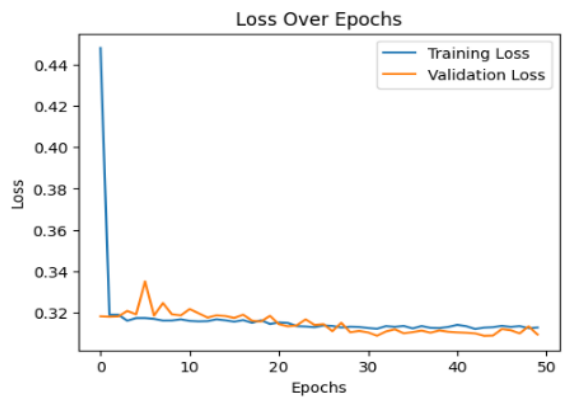

(b) Loss curve

Fig. 4. The learning curves for proposed MLP model.

### C. Baseline Models

To assess RELIANCE's performance thoroughly, we conduct a comparative analysis with the state-of-the-art baselines, which were performed on Fake News dataset:

● Ghanem et al. [11]: They have introduced a methodology that integrates lexical, word embeddings, and n-gram features for identifying the stance in fake news. Given a news title-article pair, the system focuses on assessing the relevance between the article and the title.

● Singh et al. [12]: In their study, Singh et al. employed various machine learning methods to detect fake news, utilizing the LIWC [16] (Linguistic Analysis and Word Count-based approach).

● **Ahmed et al.** [13]: In this paper, the authors to perform news credibility evaluation have introduce a hybrid model based on LR and n-gram analysis. They have used Term TF-IDF as feature extraction technique.

● **Ruchansky et al.** [15]: In this study, they introduced a model that integrates several key features of text document (such as the text of news document, news source, etc.) for a more precise and automated prediction. Specifically, their approach incorporates the behaviors of both users and articles, along with the collective behavior of users who disseminate fake news.

TABLE III. THE PARAMETER SETTINGS OF THE PROPOSED BiLSTM MODEL.

| Parameter | Setting |
|---|---|
| Number of epochs | 25 |
| Optimizer | Adam |
| Learning rate | 3e−4 |
| Loss function | Binary crossentropy |
| Activation | softmax |
| Early stopping | Applied on validation accuracy |
| Patient value | 5 |
| Dropout | 0.2 |
| Batch size | 32 |
| Cross validation | 10 fold |

TABLE IV. THE PARAMETER SETTINGS OF THE PROPOSED MLP MODEL

| Parameter | Setting |
|---|---|
| Number of epochs | 50 |
| Activation | Tanh |
| Alpha | 0.0001 |
| Hidden layer sizes | 64,128,64 |
| Learning rate | Constant |
| Solver | Adam |

TABLE V. EVALUATION RESULTS (IN PERCENT) FOR PROPOSED CREDIBILITY EVALUATION MODELS

| | Precision | Recall | F1-score | Accuracy |
|---|---|---|---|---|
| BiLSTM | 71 | 77 | 74 | 73.57 |
| LR | 87 | 89 | 88 | 87.58 |
| SVM | 91 | 88 | 89 | 89.29 |
| Random Forest | 77 | 85 | 81 | 80.54 |
| Naïve Bayes | 82 | 93 | 87 | 86.00 |
| **Ensemble learning model** | **92** | **94** | **93** | **92.43** |

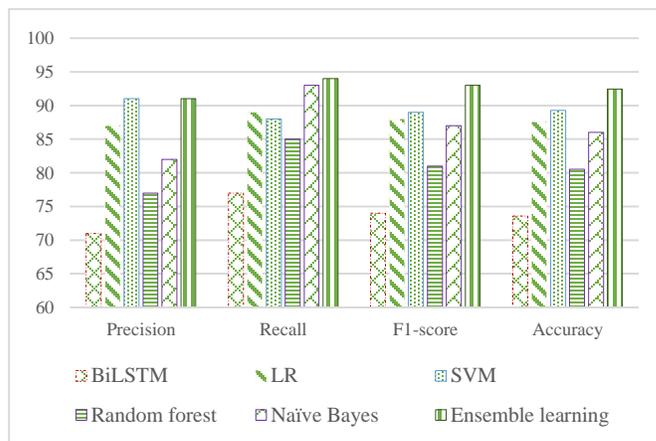

Fig. 5. Precision, recall, F1-score, and accuracy values for suggested credibility evaluation models, including 5 base models along with ensemble learning model

● **Ahmed et al.** [13]: In this paper, the authors have introduced a model for detecting fake news that employs n-gram analysis and machine learning techniques. They explore and contrast two distinct feature extraction methods along with six diverse machine classification techniques.

- **Yang et al.** [30]: They introduced a model called TI-CNN (Text and Image Information-based Convolutional Neural Network) for evaluating the credibility of news documents. Their emphasis on sensitivity analysis in their approach aimed to enhance accuracy.

Table VI presents a perspective on the performance of baseline models, along with the performance of proposed ensemble learning method by RELIANCE. Examining the data in this table reveals that RELIANCE has achieved the highest accuracy and f-measure results.

TABLE VI. COMPARING THE PERFORMANCE OF BASELINE MODELS IN CREDIBILITY EVALUATION, WHICH WERE PERFORMED ON FAKE NEWS DATASET (MISSING VALUES ARE NOT REPORTED). SIGNIFICANT VALUES ARE IN BOLD.

|  | *Precision* | *Recall* | *F1-score* | *Accuracy* |
|---|---|---|---|---|
| Ghanem et al. [11] | - | - | 48.80 | - |
| Singh et al. [12] | 86.00 | 90.00 | 88.00 | 87.00 |
| Ahmed et al. [13] | - | - | - | 89.00 |
| Ruchansky et al. [14] | - | - | 89.40 | 89.20 |
| Ahmed et al. [13] | - | - | - | 92.00 |
| Yang et al. [30] | 92.20 | 92.77 | 92.10 | - |
| RELIANCE (proposed method) | **92.49** | **93.88** | **92.75** | **92.43** |

### D. Discussion

The ensemble learning approach proposed by RELIANCE in this study, integrating five diverse base models (BiLSTMs, SVM, logistic regression, naive Bayes, and random forest) has demonstrated remarkable efficacy in the credibility evaluation of news documents. The integration of these base models through a MLP as the meta-model has yielded superior accuracy when compared to the performance of individual base models.

The robustness and versatility of the ensemble learning model stems from its ability to capture and harness the distinctive strengths of each base model, compensating for their individual limitations. The inherent diversity among the base models, arising from variations in their underlying algorithms and processing methodologies, contributes to a more comprehensive and nuanced understanding of the complex features present in news documents.

The significant improvement in accuracy achieved by the ensemble approach highlights the synergy of diverse models working collaboratively. The ensemble not only mitigates the risk of overfitting but also enhances the generalization capacity of the model, making it adept at handling the intricacies and nuances inherent in real-world news datasets. Moreover, the ensemble's outperformance suggests that the collaborative decision-making process employed by diverse models leads to a more robust and reliable system for news credibility assessment. The meta-model's ability to discern and integrate the collective insights of the base models results in a more discerning and accurate evaluation, overcoming the limitations associated with relying on any single model. Comparing the performance of RELIANCE by the state of the art studies that have been performed credibility evaluation using Fake News dataset, as shown in Table VI, clearly demonstrates the superiority of RELIANCE in credibility evaluation. Fig. 6 depicts the comparison of the accuracy of the baseline models, and our proposed method (RELIANCE) for news credibility evaluation.

Lets' to consider the main contributions of our study that was stated in Introduction:

**Contribution 1** (introducing five distinct diverse methods for news credibility evaluation): The introduction of five distinct diverse methods for news credibility evaluation serves as a foundational contribution to the field. Our choice of these methods, encompassing SVM, logistic regression, naive Bayes, random forest, and BiLSTMs, was driven by the aim to offer a diverse set of approaches, each capturing different aspects of the complex nature of documents. The subsequent utilization of these methods as base models in our ensemble learning framework allowed for a comprehensive exploration of the various features and patterns inherent in news content. Our findings highlight the importance of methodological diversity in tackling the multifaceted challenges of credibility assessment, offering valuable insights into the nuanced aspects of information reliability.

**Contribution 2** (enhancing credibility evaluation accuracy through ensemble learning): The primary objective of our study was to enhance the accuracy of news credibility evaluation, and the incorporation of ensemble learning stands out as a pivotal contribution in achieving this goal. By integrating the diverse perspectives and strengths of the five base models through a MLP as the meta-model, we witnessed a discernible enhancement in accuracy compared to individual base models. The ensemble learning framework effectively leveraged the complementary strengths of each method, mitigating the limitations associated with any singular approach. This not only resulted in a more robust and accurate credibility assessment but also highlighted the potential of ensemble learning as a powerful tool in information reliability studies.

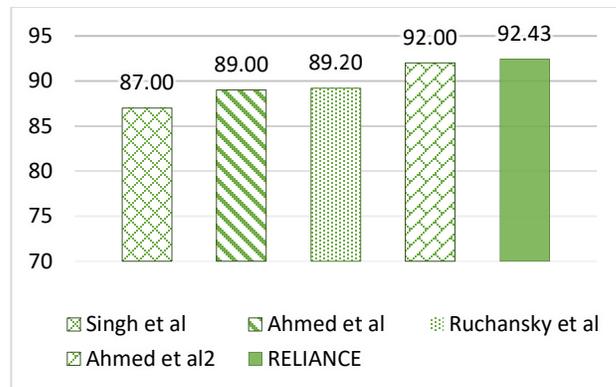

Fig. 6. Comparing the accuracy of baseline models (those that reported the accuracy values) in news credibility evaluation with the proposed model (RELIANCE)

### V. CONCLUSION

In the current era of information overload, accurately assessing the credibility of news sources is crucial for informed decision-making and effective crisis management.

To address this challenge, we propose *RELIANCE* (*Reliable Ensemble Learning for Information and News Credibility Evaluation*), an ensemble learning approach that combines the strengths of five individual models for news credibility evaluation. The base models include Support Vector Machines, naïve Bayes classifiers, logistic regression models, random forests, and Bidirectional Long Short-Term Memory Networks. These models are individually trained to extract relevant features from news documents and classify their credibility. To further enhance the overall accuracy, RELIANCE employs a multi-layer perceptron as a meta-model, which integrates the predictions of each base model (using stacking) and produces a more refined credibility assessment. Comparative experiments with baseline models demonstrate that RELIANCE significantly outperforms existing algorithms in evaluating the credibility of news documents. It provides a robust framework for identifying trustworthy news sources, offering real-world applications that empower users, journalists, and fact-checkers with a resilient tool against misinformation in the digital era.